\documentclass[aps,prd,groupedaddress,showpacs,twocolumn]{revtex4}
\usepackage[parfill]{parskip}    % Activate to begin paragraphs with an empty line rather than an indent
\usepackage{graphicx}
\usepackage{amssymb}
\usepackage{epstopdf}
\usepackage{color}
\usepackage{float}
\usepackage{ulem}
\usepackage{graphicx}
\usepackage{pdfpages}
\usepackage[colorlinks=true,citecolor=blue,urlcolor=magenta,breaklinks]{hyperref}
\begin{document}

\title{Gravitational cracking and complexity in the framework of gravitational decoupling}

\author{E. Contreras }
\email{econtreras@usfq.edu.ec}
\affiliation{Departamento de F\'isica, Colegio de Ciencias e Ingenier\'ia, Universidad San Francisco de Quito,  Quito 170901, Ecuador\\}

\author{E. Fuenmayor}
\email{ernesto.fuenmayor@ciens.ucv.ve}
\affiliation{Centro de F\'isica Te\'orica y Computacional,\\ Escuela de F\'isica, Facultad de Ciencias, Universidad Central de Venezuela, Caracas 1050, Venezuela}

%%%%%%%%%%%%%%%%%%%%%%%%%%%%%%%%%%
\begin{abstract}
In this work we analyse the stability of self gravitating spheres in the context of gravitational cracking. Besides exploring the role played by the anisotropy in the occurrence of cracking, we also study the effect of the complexity factor recently introduced by L. Herrera in Phys. Rev. D 97, 044010 (2018). The models under study correspond to anisotropic solutions obtained in the framework of the Gravitational Decoupling. The effect that the variation of the decoupling parameter and the compactness of the source have on the behaviour of the radial force is studied in detail.
\end{abstract}
%%%%%%%%%%%%%%%%%%%%%%%%%%%%%%%%%%%%%%%%%%

\maketitle

%%%%%%%%%%%%%%%%%%%%%%%%%%%%%%%%%%%%
\section{Introduction}\label{intro}

In 1992 the idea of cracking produced in a spherical fluid distribution, was first raised by L. Herrera and then fine-tuned in later works \cite{LHCracking,EFCracking,VVCracking,VV2Cracking}. The concept of cracking was introduced to describe the behavior of a fluid distribution just after its departure from equilibrium, when total nonvanishing radial forces of different signs appear within the system. We say that there is a cracking, if just after the fluid departures from equilibrium, whenever this radial force is directed inward in the inner part of the sphere and reverses its sign beyond some value of the radial coordinate. In the opposite case, when the force is directed outward in the inner part and changes sign in the outer part, we shall say that there is an overturning. Further developments on this issue may be found in \cite{LNCracking,LN2Cracking,LN3Cracking,SMCracking,4p}. As should be clear at this point, the concept of cracking is closely related to the problem of structure formation of the compact object, only at time scales that are smaller than, or at most, equal to, the hydrostatic time scale \cite{Estructura,Estructura1,Estructura2}. What we do is to take a ``snapshot'' just after the system leaves the equilibrium. To find out whether or not the system will return to the state of equilibrium afterward, is out of the scope of our analysis, and would require an integration of the evolution equations for a finite period of time, greater than the hydrostatic time. However, all this having been said, it is clear that the occurrence of cracking would drastically affect the future structure and evolution of the compact object. In \cite{EFCracking} it was shown that cracking results only if, in the process of perturbation leading to departure from equilibrium, the local anisotropy is perturbed suggesting  that fluctuations of local anisotropy may be crucial in the occurrence of cracking. We know that even small deviations from local isotropy may lead to drastic changes in the evolution of the system as it can be seen by the study of the dynamical stability of a locally anisotropic fluid \cite{CHH}, so the appearance of cracking in the initial trend of evolution of the system constitutes a real possibility. 

The number of physical processes giving rise to deviations from local isotropy that are plausible in real scenarios of stellar evolution and astrophysics in the high density regime is very large (see Refs. \cite{LHreport,Orlenis,LHMO,LHOA,LHNS,Aniso2,Aniso3,Aniso4,Aniso5,Aniso6,Aniso7}, and references therein for an extensive discussion on this point). Among all possible sources of anisotropy (see \cite{LHreport} for a discussion on this point), let us mention two which might be particularly related to our primary interest. The first one is the intense magnetic field observed in compact objects such as white dwarfs, neutron stars, or magnetized strange quark stars (see, for this point, Refs. \cite{Mag1,Mag2,Mag3,Mag4,Mag5}). It is a well established fact that a magnetic field acting on a Fermi gas produces pressure anisotropy \cite{Mag6,Mag7,Mag8,Mag9}. In a sense, the magnetic field can be addressed as a fluid anisotropy. 
Another source of anisotropy expected to be present in highly dense matter, is the viscosity (see \cite{V1,V2,V3,V4,V5,V6,V7}). As we mentioned earlier, at this point it is worth noticing that we are not concerned by how small the resulting anisotropy produced by any method might be, since the occurrence of cracking may happen even for slight deviations from isotropy.\\

Besides considering anisotropy as a quantity playing a fundamental role in the appearance of cracking, we could explore it in terms of another physical quantity involving both the anisotropy and the gradients in the density, namely the complexity factor. Many studies have devoted time and effort towards a rigorous definition of the degree of complexity of a system, so far, most of them resort to concepts such as information and entropy, and are based on the intuitive idea that complexity should measure \cite{C1,C2,C3,C4,C5,C6,C7,C8,C9,C10,C11}. In physics, the notion of complexity starts by considering a perfect crystal (that is completely ordered and therefore it has low information content) and the isolated ideal gas (that is completely disordered so it has maximum information), as examples of simplest models and therefore as systems with zero complexity. Attempts have been made to introduce other elements into the notion of complexity in the sense that the definition is better representative. The concept of ``disequilibrium'' was introduced in \cite{C7}, which measures the ``distance'' from the equiprobable distribution of the accessible states of the system. Therefore, this ``distance'' (disequilibrium) would be maximum for a perfect crystal, since it is far from an equidistribution among the accessible states, whereas it would be zero for the ideal gas. Then, it is established a compromise between the concepts of ``disequilibrium'' and information by defining the complexity through a quantity that is a product of these two concepts. In doing so, one ensures that the complexity vanishes for, both, the perfect crystal and the ideal gas. Also, several attempts have been made to define complexity in the context of self-gravitating systems in the framework of general relativity, although they present some aspects that are not entirely satisfactory \cite{C11,C12,C13,C14,C15}. Few years ago, L. Herrera \cite{complex1} (see \cite{complex2, complex2b, complex3,complex4} for recent developments) raised a new definition of complexity, for static and spherically symmetric self–gravitating systems, based on a quantity, that appears in the orthogonal splitting of the Riemann tensor. The proposal focuses on the fact that one of the simplest systems can be represented by an homogeneous fluid with isotropic pressure. Assuming this, a natural definition of a vanishing complexity system, and the very definition of complexity emerge in the theory of self–gravitating compact objects. Furthermore, the complexity factor allows one to obtain a kind of equivalence class of solutions
with the same complexity. In this respect, it should be interesting to study not only the influence of the complexity on cracking but on solutions belonging to the same equivalence class.\\

Setting some value for the complexity factor (for example, a system with vanishing complexity), this works like an equation of state that leads to close the Einstein's field equations. However, finding analytical solutions for such a set of equations could be a difficult (if not impossible) task. In this regard, a direct way of seeking for new solutions is by using the now well-known Gravitational Decoupling (GD) formalism \cite{ovalle2017} by the Minimal Geometric Deformation (MGD) approach (for implementation in $3+1$ and $2+1$ dimensional spacetimes see 
\cite{Ovalle:2016pwp,ovalle2018,ovalle2018bis,estrada2018,ovalle2018a,lasheras2018,estrada,rincon2018,ovalleplb,tello2019,lh2019,estrada2019,gabbanelli2019,ovalle2019a,sudipta2019,linares2019,leon2019,casadioyo,tello2019c,arias2020,abellan20,tello20,rincon20a,jorgeLibro,Abellan:2020dze,Ovalle:2020kpd,Contreras:2021yxe,Heras:2021xxz} and references there in) which not only allows to use a well--known solution as a {\it seed} to generate new ones but reduce the problem to solve a set of simpler differential equations. Indeed, in Ref. \cite{casadioyo} it has been obtained new interior solutions in the framework of GD and this is precisely the work in which we base our present study. It should be emphasized that the use of GD to study the effect of anisotropy in the occurrence of cracking is not mandatory. However, given that the use of GD is both simple and straightforward we have decided to implement it as the generating mechanism of anisotropic solutions.\\

This work is organized as follows. In the next section we summarize the basics for the definition of the complexity factor and the vanishing complexity condition. In Sect. \ref{mgd}, we study the basic equations of general relativity as well as the MGD formalism. Sect. \ref{cracking} is dedicated to the study of cracking for some particular models. Finally, the last section is devoted to final remarks and conclusions.

\section{The Complexity Factor}\label{complexity}

This section is dedicated to summarizing the essential aspects of the definition for the complexity factor introduced by L. Herrera \cite{complex1}. \\
It can be shown \cite{LH-C2} that the Riemann tensor can be expressed through the tensors 
\begin{eqnarray} \label{OS1}
Y_{\alpha \beta}&=& R_{\alpha\gamma\beta\delta} u^{\gamma} u^{\delta}\\
Z_{\alpha \beta}&=& ^\ast R_{\alpha\gamma\beta\delta} u^{\gamma} u^{\delta}\\
X_{\alpha \beta}&=& ^\ast R^{\ast}_{\alpha\gamma\beta\delta} u^{\gamma} u^{\delta}
\end{eqnarray}
in what is called the orthogonal splitting of the Riemann tensor \cite{Bel}. Here $\ast$ denotes the dual tensor, i.e. $R^{\ast}_{\alpha\beta\gamma\delta}=\frac12 \eta_{\epsilon\mu\gamma\delta} R_{\alpha\beta}^{\quad\epsilon\mu}$, $\eta_{\mu\nu\lambda\rho}$ corresponds to the Levi--Civita tensor and $u^{\mu}$ is a four velocity such that $u^{\mu}u_{\mu}=-1$.  Let us consider Einstein equations in the case of a spherically symmetric static anisotropic fluid, namely
\begin{eqnarray}\label{EFE}
R_{\mu\nu}-\frac{1}{2}g_{\mu\nu}R=-\kappa T_{\mu\nu},
\end{eqnarray}
with the metric defined by the line element
\begin{eqnarray}\label{metrica}
ds^{2}=e^{\nu}dt^{2}-e^{\lambda}dr^{2}-r^{2}(d\theta^{2}+\sin^{2}\theta d\phi^{2}),
\end{eqnarray}
where $\kappa=8\pi$, $T^{\mu}_{\nu}=diag(\rho,-p_{r},-p_{\perp},-p_{\perp})$ and $\nu$, $\lambda$ are functions only of the variable $r$. After some work (see \cite{complex1} for details), we can find explicit expressions for the tensors $Y_{\alpha \beta}$, $Z_{\alpha \beta}$ and $X_{\alpha \beta}$ in terms of the physical variables, getting,
\begin{eqnarray} \label{Y}
Y_{\alpha \beta} = \frac{4\pi}{3} (\rho + 3 P) h_{\alpha \beta} + 4\pi \Pi_{\alpha\beta} + E_{\alpha\beta},
\end{eqnarray}
\begin{eqnarray} \label{Z}
Z_{\alpha \beta} = 0
\end{eqnarray}
and
\begin{eqnarray} \label{X}
X_{\alpha \beta} = \frac{8\pi}{3} \rho h_{\alpha\beta} +  4\pi \Pi_{\alpha\beta} - E_{\alpha\beta},
\end{eqnarray}
with
\begin{eqnarray} \label{OS2}
\Pi^{\mu}_{\nu}=\Pi \left(s^{\mu}s_{\nu} + \frac{1}{3} h^{\mu}_{\nu} \right) ; &\quad & P=\frac{p_{r}+ 2p_{\perp}}{3}\nonumber\\
\Pi=p_{r}-p_{\perp} ; &\quad & h^{\mu}_{\nu} = \delta^{\mu}_{\nu}-u^{\mu}u_{\nu},
\end{eqnarray}
and $s_\mu$ being defined by
\begin{eqnarray} \label{s_mu}
s^{\mu}=(0, e^{-\frac{\lambda}{2}},0,0),
\end{eqnarray}
with the properties $s^{\mu}u_{\mu}=0$, $s^{\mu}s_{\mu}=-1$. $Z_{\alpha \beta}$ vanishes in the static case (see \cite{LH-C3} for details). In expressions (\ref{Y}) and (\ref{X}) $E_{\alpha\beta} = C_{\alpha\gamma\beta\delta} u^{\gamma} u^{\delta}$ (with $C_{\alpha\gamma\beta\delta}$ the components of the Weyl tensor) 
is the electric part of the Weyl tensor (in the spherically symmetric case, the magnetic part of the Weyl tensor vanishes). Observe that $E_{\alpha\beta}$ may also be written as \cite{LH-C3}, 
\begin{eqnarray} \label{E}
E_{\alpha \beta} = E \left(s_{\alpha}s_{\beta} + \frac{1}{3} h_{\alpha\beta}\right),
\end{eqnarray}
with
\begin{eqnarray} \label{E2}
E=-\frac{e^{-\lambda}}{4}\left[\nu '' + \frac{\nu '^{2} -\lambda ' \nu '}{2}- \frac{\nu ' - \lambda '}{r} + \frac{2(1-e^{\lambda})}{r^2}\right].\nonumber\\
\end{eqnarray}

From the tensors $X_{\alpha\beta}$ and $Y_{\alpha\beta}$ we can define four scalars functions, in terms of which these tensors may be written, these may be expressed as,
\begin{eqnarray} \label{XT}
X_{T} = 8\pi \rho ,
\end{eqnarray}
\begin{eqnarray} \label{XTF}
X_{TF} =  4\pi \Pi - E ,
\end{eqnarray}
\begin{eqnarray} \label{YT}
Y_{T} = 4\pi \left(\rho  + 3 p_{r} -  2 \Pi \right),
\end{eqnarray}
and
\begin{eqnarray} \label{YTF}
Y_{TF} = 4\pi \Pi + E.
\end{eqnarray}
From the above it follows that local anisotropy of pressure is determined by $X_{TF}$ and $Y_{TF}$ by
\begin{eqnarray} \label{XTF+YTF}
X_{TF} + Y_{TF} =  8 \pi \Pi ,
\end{eqnarray}
and a simple but instructive calculation performed in \cite{complex1,LH-C3} allows us to express $Y_{TF}$ in terms of the inhomogeneity of the energy density and the local anisotropy of the system like, 
\begin{eqnarray} \label{YTF2}
Y_{TF} = 8\pi \Pi - \frac{4\pi}{r^3}\int^{r}_{0} \tilde{r}^3 \rho' d\tilde{r}.
\end{eqnarray}
Also, this last result leads us to be able to write Tolman's mass as,
\begin{eqnarray} \label{m_T}
m_{T} = (m_{T})_{\Sigma}\left(\frac{r}{r_{\Sigma}}\right)^3 + r^3\int^{r_{\Sigma}}_{r} \frac{e^{( \nu + \lambda )/2}}{{\tilde{r}}} Y_{TF} d\tilde{r}.
\end{eqnarray}

Then, it is assumed that at least one of the simplest systems is represented by a homogeneous energy density distribution with isotropic pressure. For such a system the structure scalar $Y_{TF}$ vanishes. Furthermore, this single scalar function, encompasses all the modifications produced by the energy density inhomogeneity and the anisotropy of the pressure, on the active gravitational (Tolman) mass so there is a solid argument to define the complexity factor by means of this scalar. 

According to (\ref{YTF2}), the vanishing complexity factor condition, reads:
\begin{eqnarray} \label{YTF=0}
\Pi = \frac{1}{2 r^3}\int^{r}_{0} \tilde{r}^3 \rho' d\tilde{r}.
\end{eqnarray}
(\ref{YTF=0}) may be regarded as a non–local equation of state (similar to the one proposed in \cite{LN}), so we can use it to impose a condition on the physical variables when solving the Einstein equations for a static, spherically symmetric anisotropic fluid. Accordingly, if we impose the condition $Y_{TF}=0$ we shall still need another condition in order to solve the system.

\section{Field equations and gravitational decoupling}\label{mgd}

In this section we briefly review the main aspects on GD by MGD (a more detailed discussion can be found in \cite{ovalle2017}, for example). Let us start by considering the Einstein field equations (\ref{EFE})
%\begin{equation}\label{EFE}
 %   R_{\mu\nu} - \frac{1}{2}R g_{\mu\nu} = -\kappa T_{\mu\nu}^{(tot)}\;,
%\end{equation}
but now we assume that they are  sourced by certain $T_{\mu\nu}^{(tot)}$  which can be decomposed as
\begin{equation}\label{energy-momentum}
    T_{\mu\nu}^{(tot)} = T^{(m)}_{\mu\nu} + \alpha\Theta_{\mu\nu}\;.
\end{equation}
In the above equation $T^{(m)}_{\mu\nu}$ corresponds to the matter of a known solution of Einstein's field equations, namely the {\it seed} sector,  and the $\Theta_{\mu\nu}$ term describes an extra source, that is coupled by means of the parameter $\alpha$. It is essential to point out that the additional term $\alpha \Theta_{\mu \nu}$ is not considered a perturbation, i.e., the coupling parameter $\alpha$ could indeed be larger than unity. Thus, such coupling is introduced in order to control the effect of the unknown anisotropic source. Since the Einstein tensor satisfies the Bianchi identity, the total energy momentum tensor $T_{\mu\nu}^{(tot)}$ satisfies
\begin{equation}\label{divergencia-cero-total}
    \nabla_{\mu} T^{\mu\nu(tot)} = 0\;,
\end{equation}
which can be interpreted as a conservation equation. It is important to point out that, as this equation is fulfilled and given that for the matter sector (which is also a solution to Einstein's equations) we have $\nabla_\mu T^{\mu\nu(m)} = 0$, then the following condition necessarily must be satisfied
\begin{equation}\label{divergencia-cero-theta}
    \nabla_\mu \Theta^{\mu\nu} = 0\;.
\end{equation}
In this sense, there is no exchange of energy-momentum tensor between the matter solution and the anisotropic $\Theta^{\mu\nu}$ source and henceforth the interaction is purely gravitational.\\

From now on, we shall consider a static, spherically symmetric space-time with line element given by (\ref{metrica}) %parameterized as (\ref{gmn})
%\begin{equation}\label{metrica}
 %   ds^2 = e^{\nu}dt^2 - e^{\lambda} dr^2 - r^2(d\theta^2 + \sin^2{\!\theta}\,d\phi^2)\,, 
%\end{equation}
%where the metric functions, $\nu=\nu(r)$ and $\lambda=\lambda(r)$, depend only on the radial coordinate.
from where the Einstein equations (\ref{EFE}) read
\begin{eqnarray}
    \kappa\! \left(\rho^{(m)} + \alpha\Theta^0_0\right) &=& \frac{1}{r^2} +
        e^{-\lambda}\!\left(\frac{\lambda'}{r} - \frac{1}{r^2}\right)\!,\label{mgd05}\\
    \kappa\! \left(p_r^{(m)} - \alpha\Theta^1_1\right) &=& -\frac{1}{r^2} +
        e^{-\lambda}\!\left(\frac{\nu'}{r} + \frac{1}{r^2}\right)\!,\label{mgd06}\\
    \kappa\! \left(p_{\perp}^{(m)} - \alpha\Theta^2_2\right) &=& \frac{1}{4}e^{-\lambda}\!\left(2\nu'' + {\nu'}^2 - \lambda'\nu' + 2\frac{\nu'-\lambda'}{r} \right)\!.\nonumber\\\label{mgd07}
\end{eqnarray}
%where $\kappa=8\pi$ and the primes denote derivatives with respect to radial coordinate. Also, we are assuming geometric units $c=G=1$ as usual.
The left hand side of these equations can be related with the effective or total quantities
\begin{eqnarray}
    \rho &=& \rho^{(m)} + \alpha\Theta_{0}^{0} \;,\label{mgd07a}\\
    p_{r} &=& p_r^{(m)}-\alpha\Theta_{1}^{1}   \;,\label{mgd07b}\\
    p_{\perp} &=& p_{\perp}^{(m)} -\alpha\Theta_{2}^{2} \;.\label{mgd07c}
\end{eqnarray}
Because, in general, $\Theta^1_1 \neq \Theta^2_2$, we find that the system represents an anisotropic fluid. It is important to mention that although the decomposition (\ref{energy-momentum}) seems as a simple separation of the constituents of the matter sector, given the non-linearity of Einstein’s equations, such a decomposition does not lead to a decoupling of two set of equations, one for each source involved. However, contrary to the broadly belief, the decoupling is possible in the context of MGD. To apply the MGD-decoupling scheme we introduce a geometric deformation in the metric functions given by 
\begin{eqnarray}
    \nu\;\; &\longrightarrow &\;\; \xi + \alpha g,\label{comp-temporal} \\
    e^{-\lambda}\;\; &\longrightarrow &\;\; e^{-\mu} + \alpha f\;,\label{comp-radial}
\end{eqnarray}
where $\{f,g\}$ are the so-called decoupling functions and $\alpha$ is the same free parameter that ``controls'' the deformation. It is worth mentioning that although a general treatment considering deformation in both components of the metric is possible, in this work we shall concentrate in the particular case $g = 0$ and $f \ne 0$. In this case, $f$ is the so-called deformation function that depends only on the radial coordinate. Now, replacing (\ref{comp-temporal}) and (\ref{comp-radial}) in the system (\ref{mgd05}-\ref{mgd07}), we are able to split the complete set of differential equations into two subsets. Doing so, we obtain two sets of differential equations: one describing a seed sector sourced by the conserved energy-momentum tensor of the matter sector, $T_{\mu\nu}^{(m)}$, more precisely, 
    \begin{eqnarray}
    \kappa\rho^{(m)} &=& \frac{1}{r^2} +
        e^{-\mu}\!\left(\frac{\mu'}{r} - \frac{1}{r^2}\right)\!,\label{mgd13}
        \\
    \kappa p_r^{(m)} &=& -\frac{1}{r^2} +
        e^{-\mu}\!\left(\frac{\nu'}{r} + \frac{1}{r^2}\right)\!,\label{mgd14}
        \\
    \kappa p_{\perp}^{(m)} &=& \frac{1}{4}e^{-\mu} \!\! \left( 2\nu'' + {\nu'}^2 
    - \mu' \nu' + 2\frac{\nu'-\mu'}{r} \right) \!,\nonumber\\\label{mgd15}
\end{eqnarray}
and the other set corresponding to quasi-Einstein field equations sourced by $\Theta_{\mu\nu}$
 \begin{eqnarray}
    \kappa\Theta^0_0 &=& -\frac{f}{r^2} -\frac{f'}{r}\,,\label{mgd16}\\
    \kappa \Theta^1_1 &=& -f
        \left(\frac{\nu'}{r} + \frac{1}{r^2}\right)\!,\label{mgd17}\\
    \kappa \Theta^2_2 &=& -\frac{f}{4} \left( 2\nu'' + 
    {\nu'}^2 + 2\frac{\nu'}{r}\right)
    %\nonumber\\
    %& & \hspace{.5cm} 
    -\frac{f'}{4} \left( \nu' + \frac{2}{r} \right)\!.\nonumber\\\label{mgd18}
    \end{eqnarray}
As we have seen, the components of $\Theta_{\mu\nu}$ satisfy the conservation equation $\nabla_\mu \Theta^{\mu}_{\nu}=0$, given by
\begin{eqnarray}\label{consthe}
    \Theta'^1_{1} 
     - \frac{\nu'}{2} (\Theta^0_0 - \Theta^1_1)      - \frac{2}{r}(\Theta^2_2 - \Theta^1_1)=0,
\end{eqnarray}
which is a linear combination of Eqs. (\ref{mgd16}), (\ref{mgd17}) and (\ref{mgd18}). Although the quasi–Einstein equations differ from Einstein equations, the expression given in (\ref{consthe}) is completely analogous to the anisotropic Tolman-Opphenheimer-Volkoff
(TOV) equation.

It is worth mentioning that in our case we demand for the exterior the Schwarzschild solution. So, outside of
the fluid distribution, the spacetime of the stellar model is given by,
\begin{eqnarray}\label{sol04a}
    ds^2 = \left(1 - \frac{2M}{r}\right)dt^2 &-& \left(1 - \frac{2M}{r}\right)^{-1} dr^2 +\nonumber\\ 
    &-& r^2 d\theta^2 - r^2\sin^2{\!\theta}\,d\phi^2\,.\nonumber\\
%    (d\theta^2 + \sin^2{\!\theta}\,d\phi^2)\,. 
\end{eqnarray}
Then, in order to match smoothly the interior metric with the outside one above on the boundary surface $\Sigma$, we require
\begin{eqnarray}
    e^{\nu}\Big|_{\Sigma} &=& \left(1 - \frac{2M}{r}\right)\Bigg|_{\Sigma}\,,\label{mgd11a}\\
    e^{\lambda}\Big|_{\Sigma} &=& \left(1 - \frac{2M}{r}\right)^{-1}\Bigg|_{\Sigma}\,,\label{mgd11b}\\
     p_r(r) \Big|_{\Sigma} &=& \;0\,,\label{mgd11c}
\end{eqnarray}
which corresponds to the continuity of the first and second fundamental form across that surface. From now on, subscript $\Sigma$ indicates that the quantity is evaluated on the boundary surface $\Sigma$, and (\ref{mgd11c}) expresses the fact that the radial pressure must vanish at the boundary surface as usual. Then, the matching condition leads to the extra information required to completely solve the system. Note that the condition on the radial pressure leads to
\begin{eqnarray}
     p^{(m)}_r(r_\Sigma) - \alpha \Theta^1_1 (r_\Sigma) = \;0\,,\label{pradial}
\end{eqnarray}
so we see that, if the original matter fluid match smoothly with
the Schwarzschild solution, i.e, $p^{(m)}_{r}(r_\Sigma)=0$, Eq. (\ref{pradial}) can be satisfied by demanding that $\Theta^1_1 \sim p$. Of course, is known that the simpler way to satisfy the requirement on the radial pressure is assuming the so–called mimic constraint \cite{ovalle2017} for the pressure, namely
\begin{eqnarray}
     p_r^{(m)} = \Theta^1_1 \label{mimic}
\end{eqnarray}
in the interior of the star. Remarkably, this condition leads to an algebraic equation for $f$ such that, in principle, any solution can be extended with this constraint. Another possibility, that has been reported recently \cite{abellan20} consists of implementing physical requirements on the anisotropy function induced by the decoupling sector, $\Theta^2_2 - \Theta^1_1$, so you can find an anisotropic solution assuming a regularity condition on the anysotropy function of the decoupling sector.\\

In this work, we shall use anisotropic solutions previously obtained in \cite{casadioyo} based on different equations of state. The fist solution was obtained by assuming the isotropization through GD of an ansotropic solution, so that the equation of state corresponds to $p_{r}-p_{\perp}=0$ for the total solution which leads to
\begin{eqnarray}\label{f0}
\frac{f'}{32\pi}\left(\nu'+\frac{2}{r}\right)&+&\frac{f}{32\pi}\left(2\nu''+\nu'^{2}-\frac{2\nu '}{r}-\frac{4}{r^{2}}\right)\nonumber\\
&&+p_{r}^{(m)}-p_{\perp}^{(m)}=0.
\end{eqnarray}

The other models we shall consider were obtained using
the complexity factor in the framework of GD to provide an equation of state which allows to close the system as follows. In \cite{casadioyo}  was demonstrated that in the framework of GD the complexity factor behaves as an additive quantity. More precisely, let $Y_{TF}$ be the complexity associated to $T_{\mu\nu}^{(tot)}$, then we have
\begin{eqnarray}
Y_{TF}=Y_{TF}^{m}+Y^{\theta}_{TF},
\end{eqnarray}
where $Y_{TF}^{m}$ and $Y^{\theta}_{TF}$ corresponds to the complexity for the seed and the $\theta$--sectors, respectively. In this respect, we can construct solutions through GD with the same complexity factor than the seed sector by imposing $Y_{TF}^{ \theta}=0$, which leads to \cite{casadioyo}
\begin{eqnarray}\label{f1}
f'\left(\nu'+\frac{4}{r}\right)+
f\left(2\nu''+\nu'^{2}-\frac{2\nu'}{r}-\frac{8}{r^{2}}\right)=0 .
\end{eqnarray}
Similarly, we can construct a family of solutions with a vanishing total complexity which produces \cite{casadioyo}
\begin{eqnarray}\label{f2}
f'\left(\nu'+\frac{4}{r}\right)+
f\left(2\nu''+\nu'^{2}-\frac{2\nu'}{r}-\frac{8}{r^{2}}\right)+Y_{TF}=0,\nonumber\\
\end{eqnarray}
where we need to provide some particular seed solution to obtain $Y_{TF}$.\\

For later discussions it is useful to end this section by writing the explicit form of the radial component of the conservation law for the total fluid, namely equation (\ref{divergencia-cero-total}),
\begin{eqnarray}\label{TOV}
p_{r}' = - \frac{\nu '}{2} (\rho + p_{r}) + \frac{2(p_{\perp} - p_{r})}{r},
\end{eqnarray}
which is nothing other than the hydrostatic equilibrium (TOV) equation for an anisotropic fluid. It is worth noticing that on the one hand, in Eq. (\ref{TOV}) the pressure gradient is balanced by a gravitational  term  (that  has  the  derivative  of  the  metric  variable $\nu$ present)  and  a  term  that  includes  the  local anisotropy  distribution. On the other hand, the TOV has dimensions of force per unit volume so, the quantity
\begin{eqnarray}
\mathcal{R}\equiv p_{r}' + \frac{\nu '}{2} (\rho + p_{r}) - \frac{2(p_{\perp} - p_{r})}{r},
\end{eqnarray}
is the total force per unit volume on each fluid element. It is clear that when the system is in Equilibrium, $\mathcal{R}=0$. However, after perturbation, the total force is not vanishing anymore as we shall discuss in more detail in the next section.

Introducing the mass function $m(r)$ from
\begin{eqnarray}\label{funcion-masa}
e^{- \lambda}= 1 - \frac{2m}{r},
\end{eqnarray}
we can write (\ref{TOV}) as
\begin{eqnarray}\label{fuerza-radial}
\mathcal{R} & =&\frac{dp_r}{dr} + \frac{4\pi r p_{r}^{2}}{1 - 2m/r} + \frac{m p_r}{r^2 (1 - 2m/r)} + \frac{4\pi r \rho p_{r}}{1 - 2m/r}
 \nonumber\\ &+& \frac{\rho m}{r^2 (1 - 2m/r)} + \frac{2(p_{\perp} - p_{r})}{r} = 0.
\end{eqnarray}
For this, also we have used the relationship
\begin{eqnarray}\label{funcion-masa2}
m(r)= 4 \pi \int_{0}^{r} T^{0}_{0} r^2 dr,
\end{eqnarray}
easily deduced from (\ref{mgd05}) and (\ref{funcion-masa}).\\

In the next section we shall explore the consequences of perturbing the system such that the total force $\mathcal{R}$ is not vanishing anymore. As we shall see later, such a perturbation should lead to the cracking of the system.

\section{Cracking for self gravitating spheres}\label{cracking}
The gravitational cracking corresponds to the situation in which after the fluid departures from equilibrium, its radial force is directed inward in the inner part of the sphere and reverses its sign beyond some value of the radial coordinate, $r$. In this work, we assume that the perturbation is done in such a manner that the profile of the radial pressure remains the same but the rest of quantities (density, anisotropy, etc) undergo a change though the parameters of the model which leads to $\mathcal{R}\ne0$.
For example, let  $\{\alpha,\beta\}$ be the parameters of the model, the ``modified'' TOV can be written as $\tilde{R}(\alpha+\delta\alpha,\beta+\delta\beta)$ so, up to first order in perturbation we obtain
\begin{eqnarray*}
\tilde{R}=\frac{\partial\tilde{R}}{\partial\alpha}\delta\alpha+\frac{\partial\tilde{R}}{\partial\beta}\delta\beta+\mathcal{O}(\delta\alpha^{2},\delta\beta^{2}).
\end{eqnarray*}
Form the previous equation it is observed that cracking occurs whenever there is a change of sing of $\tilde{R}$ at some radius $r$ inside the body. To be more precise,  we say that cracking occurs when $\tilde{R}$ has at least one real root. Note that, if we define $\delta\beta=-\Gamma\delta\alpha$ with
$\Gamma$ a constant, the condition for cracking becomes in a condition for the existence of a $\Gamma$ such that
\begin{eqnarray*}
\Gamma=\frac{\partial\tilde{\mathcal{R}}/\partial\tilde{\alpha}|_{\beta,\alpha}}{\partial\tilde{\mathcal{R}}/\partial\tilde{\beta}|_{\beta,\alpha}}.
\end{eqnarray*}
In the next section, we shall apply the ideas developed here to particular interior solutions differentiated by means of there gravitational complexity parameter.

\subsection{Anisotropic Tolman IV model}\label{model1}
In this section we shall consider a model obtained in  \cite{casadioyo} by imposing (\ref{f0}) and assuming an anisotropic like--Tolman IV solution of Ref. \cite{ovalle2017} as a seed solution. The result reads
\begin{eqnarray}
e^{\nu}&=&B^2 \left(\frac{r^2}{A^2}+1\right)\\
e^{-\lambda}&=&
\frac{A^2+r^2}{A^2+3 r^2}+
\frac{3\alpha r^2 \left(A^2+r^2\right) \left(R^2-r^2\right)}{\chi(r) \left(A^2+3 R^2\right)}\\
\rho&=&\frac{3 \left(A^2+r^2\right)}{4 \pi  \left(A^2+3 r^2\right)^2}+
\frac{3 \alpha\xi(r) }{8 \pi \chi(r)^{2} \left(A^2+3 R^2\right)} \label{rho}\\
p_{r}&=&\frac{3 \alpha  \left(R^2-r^2\right)}{8 \pi  \chi(r)} \label{pr}\\
\Pi&=&-\frac{3 ( 1-\alpha) r^2}{8 \pi  \left(A^2+3 r^2\right)^2}\label{delta}\\
m&=&\frac{r^{3}}{A^{2}+3r^{2}}-
\frac{3\alpha r^{3}(A^{2}+r^{2})(R^{2}-r^{2})}{\chi(r)(A^{2}+3R^{2})}\label{m},
\end{eqnarray}
where
\begin{eqnarray}
\chi&=&\left(A^2+2 r^2\right) \left(A^2+3 r^2\right)\label{chi}\\
\xi&=&5 A^6 r^2 + 22 A^4 r^4 + 31 A^2 r^6 + 
 18 r^8\nonumber\\
 &&- (3 A^6 + 10 A^4 r^2 + 9 A^2 r^4 + 6 r^6) R^2 \label{xi}.
\end{eqnarray}
In the above expressions, $R$ is the radius of the star and the parameters $A^{2}$ and $B^{2}$, which ensure a smooth matching with the Schwarzschild exterior, 
are given by
\begin{eqnarray}
\frac{A^{2}}{R^{2}}&=&\frac{R-3M}{M}=\frac{1}{u}-3\label{restbeta}\\
B^{2}&=&1-3u,
\end{eqnarray}
where $u\equiv M/R$ is the compactness parameter.
The parameter $\alpha\in[0,1]$ is a dimensionless quantity (the decoupling parameter) inherited from the process of gravitational decoupling which in this context also controls the anisotropy of the system. Note that for $\alpha=0$, the system is in an anisotropic configuration and as far the value of the decoupling parameter approaches to $\alpha=1$ the anisotropy  decreases until reach its minimum value when $\alpha=1$. As a consequence, the solution corresponds to an isotropic fluid an in this regard $\alpha$ has a kind of screening effect on the anisotropy.
\\

Let us proceed to define dimensionless quantities 
\begin{eqnarray}
\beta&=&\frac{A}{R}\label{beta}\\
x&=&\frac{r}{R},
\end{eqnarray}
in terms of which we can rewrite the set $\{\rho,p_{r},\Pi,m\}$ given in (\ref{rho}), (\ref{pr}), (\ref{delta}) and (\ref{m}) as
\begin{eqnarray}
\rho&=&\frac{3}{4\pi R^{2}}
\bigg(
\frac{\beta^{2}+x^{2}}{\beta^{2}+3x^{2}}+\frac{\alpha}{2}\frac{\xi(\beta,x)}{\chi(\beta,x)^{2}(\beta^{2}+3)}
\bigg)\nonumber\\
&=&\frac{3}{4\pi R^{2}}\hat{\rho}(\beta,\alpha,x),\\
p_{r}&=&\frac{3\alpha}{8\pi R^{2}}\mathcal{F}(x),\\
\Pi&=&-\frac{3(1-\alpha)}{8\pi R^{2}}\frac{x^{2}}{(\beta^{2}+3x^{2})^{2}}\nonumber\\
&=&\frac{3(1-\alpha)}{8\pi R^{2}}\hat{\Pi}(\beta,x),\\
m&=&R\bigg(\frac{x^{3}}{\beta^{2}+3x^{2}}
-\frac{3\alpha x^{3}(\beta^{2}+x^{2})(R^{2}-x^{2})}{\chi(\beta,x)(\beta^{2}+3x^{2})}\bigg)\nonumber\\
&=&R\hat{m}(\beta,\alpha,x),
\end{eqnarray}
where
\begin{eqnarray}
\mathcal{F}(x)&=&\frac{1-x^{2}}{(\beta^{2}+2x^{2})(\beta^{2}+3x^{2})}\\
\chi(\beta,x)&=&(\beta^{2}+2x^{2})(\beta^{2}+3x^{2}),\\
\xi(\beta,x)&=&5\beta^{6}x^{2}+22\beta
^{4}x^{4}+31\beta^{2}x^{6}+18x^{8}\nonumber\\
&&-3\beta^{6}-10\beta^{4}x^{2}-9\beta^{2}x^{4}-6x^{6},
\end{eqnarray}
are obtained directly from (\ref{chi}) and (\ref{xi}). 

We now proceed to perturb the matter sector through variations of the parameters $\{\alpha, \beta\}$, namely
\begin{eqnarray}
\alpha&\to&\tilde{\alpha}=\alpha+\delta\alpha\label{t1},\\
\beta&\to&\tilde{\beta}=\beta+\delta\beta\label{t2},
\end{eqnarray}
where tilde indicates that the quantity is being perturbed. Explicitly we are leaving the functional dependence of $p_{r}$ unchanged. Of course, after perturbation, the TOV is different from zero, so the system is no longer in hydrostatic equilibrium (follow the discussions in \cite{LHCracking, EFCracking}). To be more precise, transformations (\ref{t1}) and (\ref{t2}) in (\ref{fuerza-radial}) leads to 
%\begin{eqnarray}
%\tilde{\mathcal{R}}&=&\frac{3\tilde{%\alpha}}{8\pi %R^{3}}\frac{d\mathcal{F}}{dx}
%+
%\frac{3}{4\pi %R^{3}}\frac{(\hat{\rho}+
%\frac{\alpha}{2}\mathcal{F})(\hat{m}%+\frac{3\alpha}{2}x^{2}\mathcal{F})}%{x^{2}(1-2\hat{m}/x)}\\
%&&
%-2\frac{3(1-\alpha)}{8\pi %R^{3}}\frac{\hat{\Delta}}{x}
%\end{eqnarray}

\begin{eqnarray}\label{rtilde}
\tilde{\mathcal{R}}&=&\frac{\tilde{\alpha}}{2}\frac{d\mathcal{F}}{dx}
+
\frac{(\hat{\rho}(\tilde{\beta},\tilde{\alpha},x)+
\frac{\tilde{\alpha}}{2}\mathcal{F})(\hat{m}(\tilde{\beta},\tilde{\alpha},x)+\frac{3\tilde{\alpha}}{2}x^{2}\mathcal{F})}{x^{2}(1-2\hat{m}(\tilde{\beta},\tilde{\alpha},x)/x)}\nonumber\\
&&
+(1-\tilde{\alpha})\frac{\hat{\Pi}(\tilde{\beta},x)}{x},
\end{eqnarray}
where
\begin{eqnarray}
\tilde{\mathcal{R}}\equiv\frac{4\pi R^{3}}{3}\mathcal{R}.
\end{eqnarray}
Formally, we may write
\begin{eqnarray}\label{formalmente}
&&\tilde{\mathcal{R}}(\beta+\delta\beta,\alpha+\delta\alpha,x)=
\tilde{\mathcal{R}}(\beta,\alpha,x)+
\frac{\partial\tilde{\mathcal{R}}}{\partial\tilde{\beta}}\bigg|_{\beta,\alpha}\delta\beta\nonumber\\
&&\hspace{2cm}+\frac{\partial\tilde{\mathcal{R}}}{\partial\tilde{\alpha}}\bigg|_{\beta,\alpha}\delta\alpha+\mathcal{O}(\delta\alpha^{2},\delta\beta^{2}),
\end{eqnarray}
where the first term is zero given that it corresponds to the unperturbed values of $\tilde{\mathcal{R}}$. Obviously, (\ref{formalmente}) (with partial derivatives evaluated) is equivalent to (\ref{rtilde}) up to terms of first order in the perturbation. Note that, if cracking occurs, $\tilde{\mathcal{R}}$
must have a zero in the interval $x\in(0,1)$. At this cracking point (at first order in the perturbation)
\begin{eqnarray}\label{gamma1}
\delta\beta=-\Gamma\delta\alpha, 
\end{eqnarray}
with
\begin{eqnarray}\label{gamma2}
\Gamma=\frac{\partial\tilde{\mathcal{R}}/\partial\tilde{\alpha}|_{\beta,\alpha}}{\partial\tilde{\mathcal{R}}/\partial\tilde{\beta}|_{\beta,\alpha}}.
\end{eqnarray}
Note that $\Gamma$, can be interpreted as a measure of how far the variations $\{\delta\alpha,\delta\beta\}$
deviate between them. In this regard, $\Gamma$ should be considered as a perturbation ratio.\\

\begin{figure}[ht!]
\centering
\includegraphics[scale=0.5]{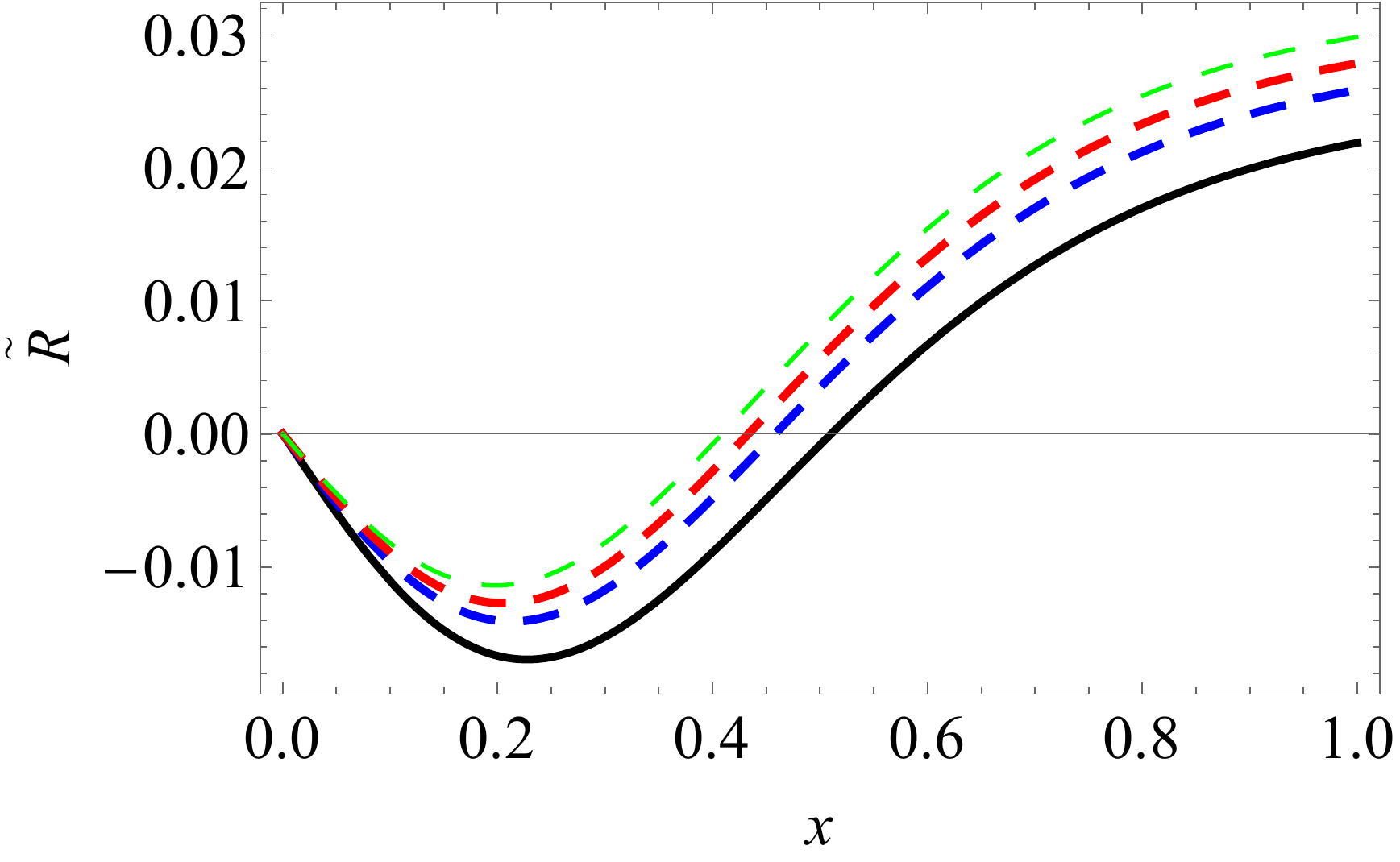}
\caption{\label{r1}
$\tilde{\mathcal{R}}$ as function of $x$, for 
$\alpha=0.6$, $\beta=1$ and $\Gamma=-1.8$ (black line), $\Gamma=-2$ (blue line), $\Gamma=-2.1$ (red line) and $\Gamma=-2.2$ (green line)
}
\end{figure}

We proceed to plot $\tilde{\mathcal{R}}$ as a function of $x$, for different values of $\alpha$, $\beta$ and $\Gamma$ in order to explore the possibility
of finding cracking or overturnings in the system. In figure \ref{r1} we show $\tilde{\mathcal{R}}$ as a function of $x$ for $\alpha=0.6$, $\beta=1$ (which, in accordance to definition (\ref{beta}) and condition (\ref{restbeta}), corresponds to a compactness parameter $u=0.25$)
and different values of $\Gamma$. We see that as the absolute value of $\Gamma$ increases cracking occurs in deeper regions of the fluid distribution which means that the size of the surface where $\tilde{R}$ has a root decreases.

\begin{figure}[ht!]
\centering
\includegraphics[scale=0.5]{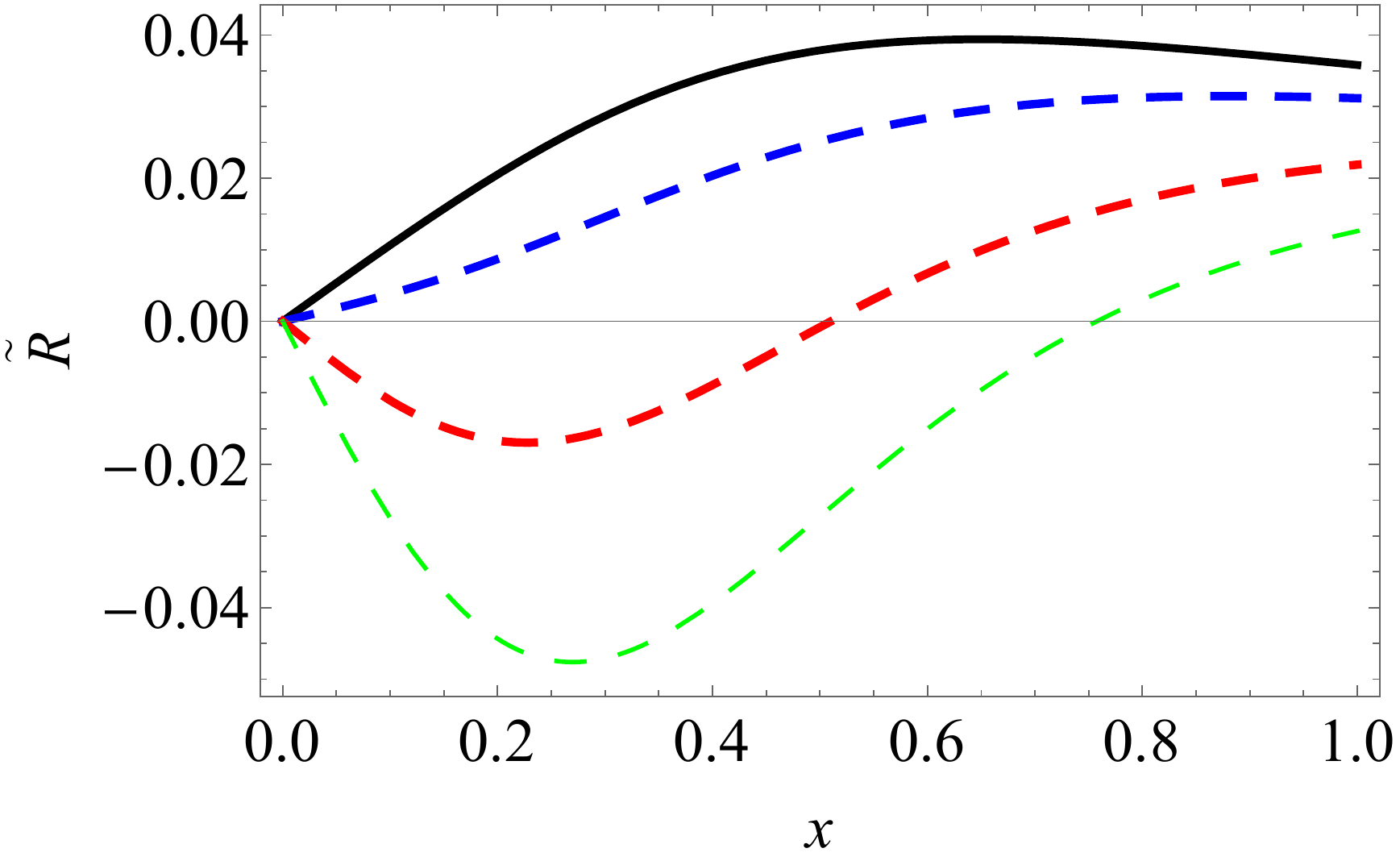}
\caption{\label{r2}
$\tilde{\mathcal{R}}$ as function of $x$, for 
$\beta=1$, $\Gamma=-1.8$ and $\alpha=0$ (black line), $\alpha=0.2$ (blue line), $\alpha=0.6$ (red line) and $\alpha=1$ (green line)
}
\end{figure}

In figure \ref{r2} we show $\tilde{\mathcal{R}}$ as a function of $x$ for $\Gamma=-1.8$, $\beta=1$
and different values of $\alpha$. 
Note that, the system goes from a configuration where there is no cracking (black line) to a situation where cracking occurs (red and green lines). It is interesting to note that the system is ``more stable" when is initially in an anisotropic configuration ($\alpha=0$)  and loss such a condition as far as we approach to the isotropic case ($\alpha=1$). More precisely, the screening effect on the anisotropy 
induced by $\alpha$ leads to
instabilities in the system and eventually to the occurrence of cracking. 
This aspect often manifests itself in the ``search'' for stability of a self-gravitating system.

\begin{figure}[ht!]
\centering
\includegraphics[scale=0.5]{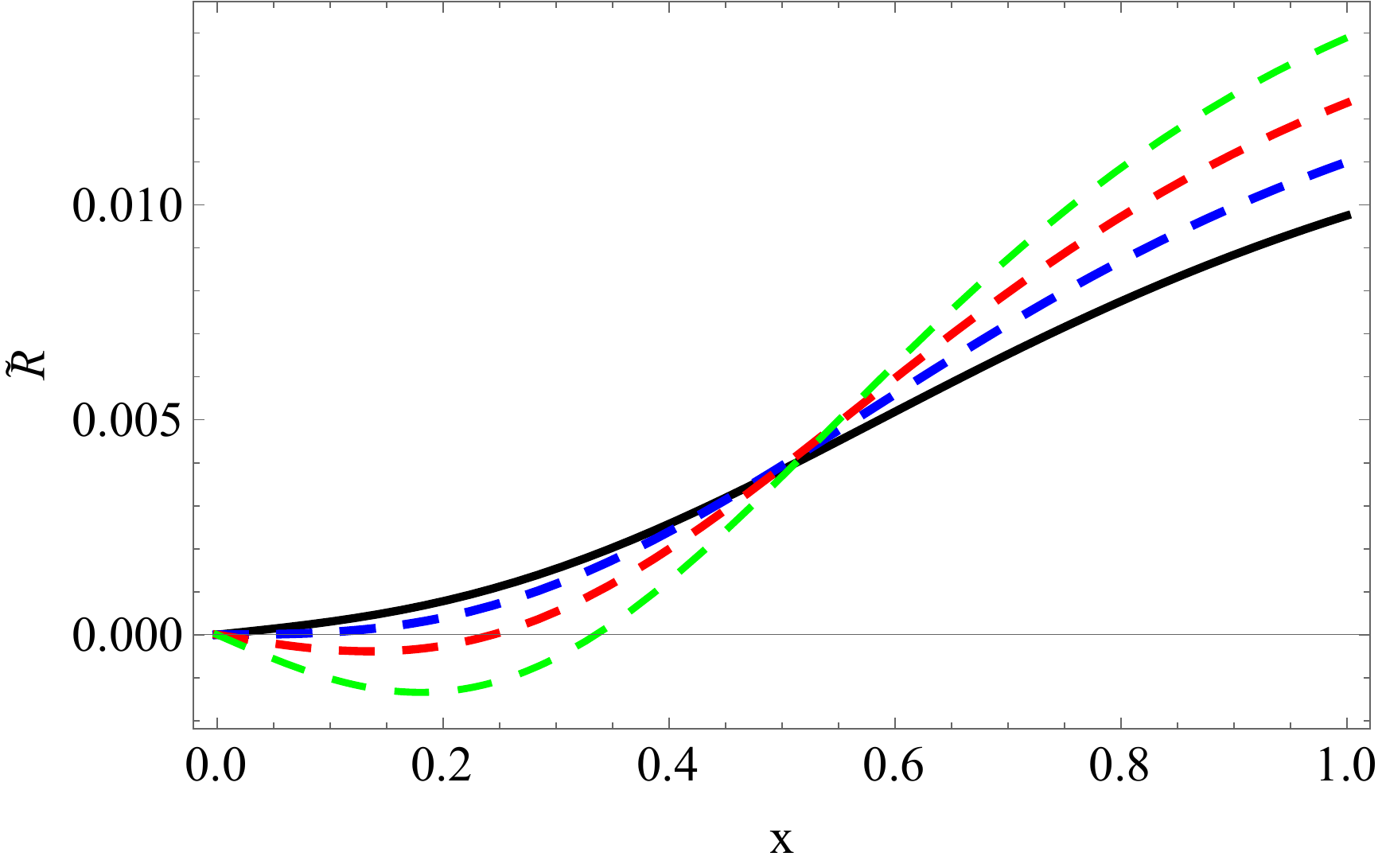}
\caption{\label{beta-var}
$\tilde{\mathcal{R}}$ as function of $x$, for $\alpha=0.6$, $\Gamma=-1.8$ and $\beta=1.6,\ (u\approx 0.18-$black line), $\beta=1.5,\ (u\approx0.19-$blue line), $\beta=1.4, \ (u\approx0.2-$red line) and $\beta=1.3,\ (u\approx 0.21-$green line),
}
\end{figure}

In figure \ref{beta-var} it is shown $\tilde{\mathcal{R}}$ as a function of $x$ for different values of $\beta$, which measure how the compactness affect the appearance of cracking. 
Note that for the values under consideration, the cracking is absent for the less compact solutions. Also, observe that the more compact the object is, the greater the value of $\mathcal{R}$ (see reference \cite{VVCracking} for details). The situation described in figures \ref{r1}, \ref{r2} and \ref{beta-var} is representative for a wide range of parameters for which there exist bounded configurations satisfying the required physical conditions.

%%%%%%%%%%%%%%%%%%%%%%%%%%%%%%%%%%%%%%%

\subsection{A family of anisotropic solutions with the same complexity}\label{model2}

In this section we shall consider the gravitational cracking on solutions with the same complexity factor. To this end we consider a family of solutions with a $\theta$--sector with vanishing complexity, namely $Y_{TF}^{\theta}=0$, and the Tolman IV as the material seed solution. In this regard, the total solution has the same complexity as the seed sector, $Y_{TF}=Y_{TF}^{(m)}$, while the decoupling sector is determined
through (\ref{f1}). The solution reads
\begin{eqnarray}
e^{\nu}&=&B^{2}
\left(
1+\frac{r^{2}}{A^{2}}
\right)\\
e^{-\lambda}&=&\frac{(C_{\alpha\ell}^{2}-r^{2})(A^{2}+r^{2})}{C_{\alpha\ell}^{2}(A^{2}+2r^{2})}+\frac{\alpha r^{2}(A^{2}+r^{2})}{\ell^{2}(2A^{2}+3r^{2})}\\
\rho&=&\rho_{0}-\frac{\alpha(6A^{4}+13A^{2}r^{2}+9r^{4})}{8\pi \ell^{2}(2A^{2}+3r^{2})^{2}}\\
p_{r}&=&p_{r0}+\frac{\alpha(A^{2}+3r^{2})}{8\pi \ell^{2}(2A^{2}+3r^{2})}\\
\Pi&=&-\frac{\alpha A^{2}r^{2}}{8\pi \ell^{2}(2A^{2}+3r^{2})^{2}}
\end{eqnarray}
with
\begin{eqnarray}
\rho_{0}&=&\frac{3A^{2}+A^{2}(3C_{\alpha\ell}^{2}+7r^{2})+2r^{2}(C_{\alpha\ell}^{2}+3r^{2})}{8\pi C_{\alpha\ell}^{2}(A^{2}+2r^{2})^{2}}\\
p_{r0}&=&\frac{C_{\alpha\ell}^{2}-A^{2}-3r^{2}}{8\pi C_{\alpha\ell}^{2}(A^{2}+2r^{2})},
\end{eqnarray}
and where 
\begin{eqnarray}
\frac{A^{2}}{R_{\alpha\ell}^{2}}&=&\frac{R_{\alpha\ell}-3M_{\alpha\ell}}{M_{\alpha\ell}}\label{A}\\
B^{2}&=&1-\frac{3M_{\alpha\ell}}{R_{\alpha\ell}}\label{B}\\
C_{\alpha\ell}&=&\frac{R_{\alpha\ell}^{3}}{M_{\alpha\ell}}\nonumber\\
&-&\frac{\alpha(A^{2}+2R_{\alpha\ell}^{2})(A^{2}+3R^{2}_{\alpha\ell})^{2}}{\alpha(A^{4}+5A^{2}R_{\alpha\ell}^{2}+6R_{\alpha\ell}^{4})+\ell^{2}(2A^{2}+3R^{2}_{\alpha\ell})}\nonumber\\
&=&\frac{R^{3}}{M}\label{Calfa}.
\end{eqnarray}
The Parameters $A$, $B$ and $C_{\alpha\ell}$ defined in
(\ref{A}), (\ref{B}) and (\ref{Calfa})
ensure the matching with the Schwarzschild exterior of a star  with radius and mass given by $R_{\alpha\ell}$ and $M_{\alpha\ell}$, respectively, and  $\ell$ is a parameter with dimensions of a length. Note that, when $\alpha\to0$,
the above solution reduces to the Tolman IV model with radius and mass $R$ and $M$, respectively. Besides, Eq. (\ref{Calfa}) represents a constraint between the parameters $\{A,R_{\alpha\ell},\alpha,\ell\}$; in particular, $\ell^{2}$ can be written as
\begin{eqnarray}\label{lcuadra}
\ell^{2}=-\frac{\alpha   \left(\beta ^4+5 \beta ^2+6\right)  \left(\beta ^2+3 \Xi ^2\right)}{3 \left(2 \beta ^2+3\right) \left(\Xi ^2-1\right)}R_{\alpha\ell}^2
\end{eqnarray}
with 
\begin{eqnarray}
A&=&\beta  R_{\alpha\ell}\\
r&=& R_{\alpha\ell} x\\
R&=& \Xi  R_{\alpha\ell}.
\end{eqnarray}
Note that the interpretation of the dimensionless quantities $\beta$ and $x$ are similar to those of the previous section. Now, the parameter $\Xi$ labels the representative of the equivalence class we are taking into account. Indeed, for $\Xi=1$ the representative corresponds to the Tolman IV solution for perfect fluids. Alternatively, $\Xi$ can be interpreted as a measure of how much the radius of the star deviates from this of the Tolman IV model in order that the solution maintains the vanishing complexity. In this regard, $\Xi$ plays the role of radius ratio.\\

A straightforward computations reveals that, after using
(\ref{lcuadra}), all the quantities
are parametrized by $\{\beta,\Xi\}$, indeed
\begin{eqnarray}
\hat{p}_{r}&=&3 f(x)\\
\hat{\rho}&=&\frac{3 \left(2 \beta ^2+3\right) \left(\Xi ^2-1\right) \left(6 \beta ^4+13 \beta ^2 x^2+9 x^4\right)}{\left(\beta ^4+5 \beta ^2+6\right) \left(\beta ^2+3 \Xi ^2\right) \left(2 \beta ^2+3 x^2\right)^2}\nonumber\\
&&+\frac{6 \beta ^4+9 \beta ^2 \left(x^2+1\right)+6 \left(x^4+x^2\right)}{\left(\beta ^2+3\right) \left(\beta ^2+2 x^2\right)^2}\\
\hat{\Pi}&=&-\frac{3 \beta ^2 \left(2 \beta ^2+3\right) \left(\Xi ^2-1\right) x^2}{  \left(\beta ^4+5 \beta ^2+6\right)  \left(\beta ^2+3 \Xi ^2\right) \left(2 \beta ^2+3 x^2\right)^2}\\
\hat{g}\hat{m}&=&
x^2 (6 (7 \beta ^4+17 \beta ^2+9) \Xi ^2
+(8 \beta ^4+7 \beta ^2-9) \beta ^2)
\nonumber\\
&&
+\beta ^2 (2 \beta ^2+3) (\beta ^2 (9 \Xi ^2+1)+2 \beta ^4+12 \Xi ^2)
\nonumber\\
&&
+3 x^4 (\beta ^2 (7 \Xi ^2-2)+\beta ^4+12 \Xi ^2-6),
\end{eqnarray}
where $\hat{p}_{r}=8 \pi  R_{\alpha\ell}^2 p_{r}$, $\hat{\rho}=8\pi R_{\alpha\ell}^{2}\rho$,
$\hat{\Pi}=8 \pi R_{\alpha\ell}^2\Pi$,
$\hat{m}=m/R_{\alpha\ell}$ and
\begin{eqnarray}
f(x)&=&
-\frac{\left(2 \beta ^2+3\right) \left(\Xi ^2-1\right) \left(\beta ^2+3 x^2\right)}{\left(\beta ^4+5 \beta ^2+6\right) \left(\beta ^2+3 \Xi ^2\right) \left(2 \beta ^2+3 x^2\right)}
\nonumber\\
&&
+\frac{1-x^2}{\left(\beta ^2+3\right) \left(\beta ^2+2 x^2\right)}
\\
\hat{g}&=&2 x^{-3}\left(\beta ^4+5 \beta ^2+6\right) \left(\beta ^2+3 \Xi ^2\right)\nonumber\\
&&\left(2 \beta ^4+7 \beta ^2 x^2+6 x^4\right).
\end{eqnarray}

\begin{figure}[ht!]
\centering
\includegraphics[scale=0.5]{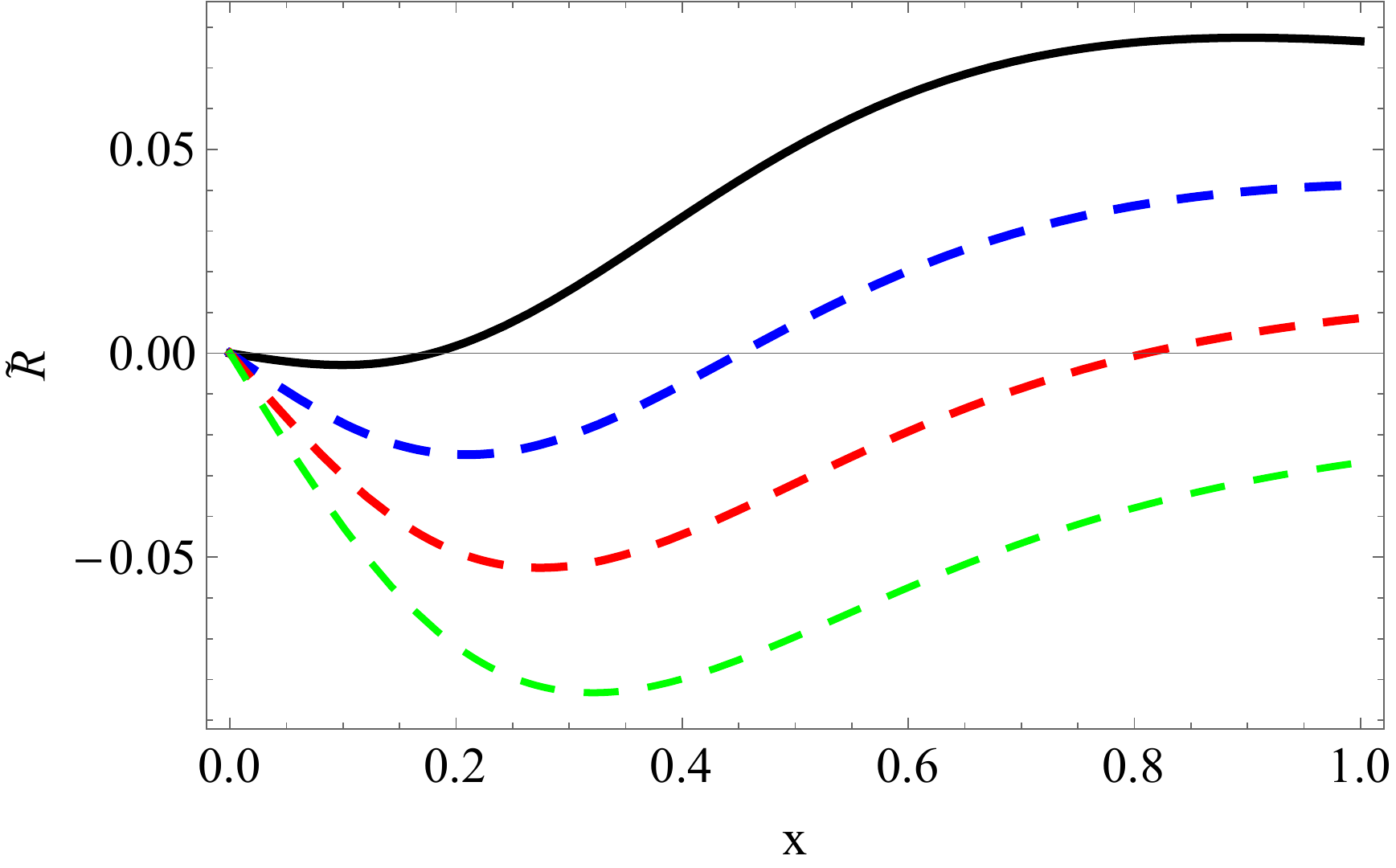}
\caption{\label{r4}
$\tilde{\mathcal{R}}$ as function of $x$, for 
$\beta=1$, $\Gamma=0.1$ and $\Xi=0.1$ (black line), $\Xi=0.3$ (blue line) and $\Xi=0.5$ (red line) and $\Xi=1$ (green line)
}
\end{figure}

\begin{figure}[ht!]
\centering
\includegraphics[scale=0.5]{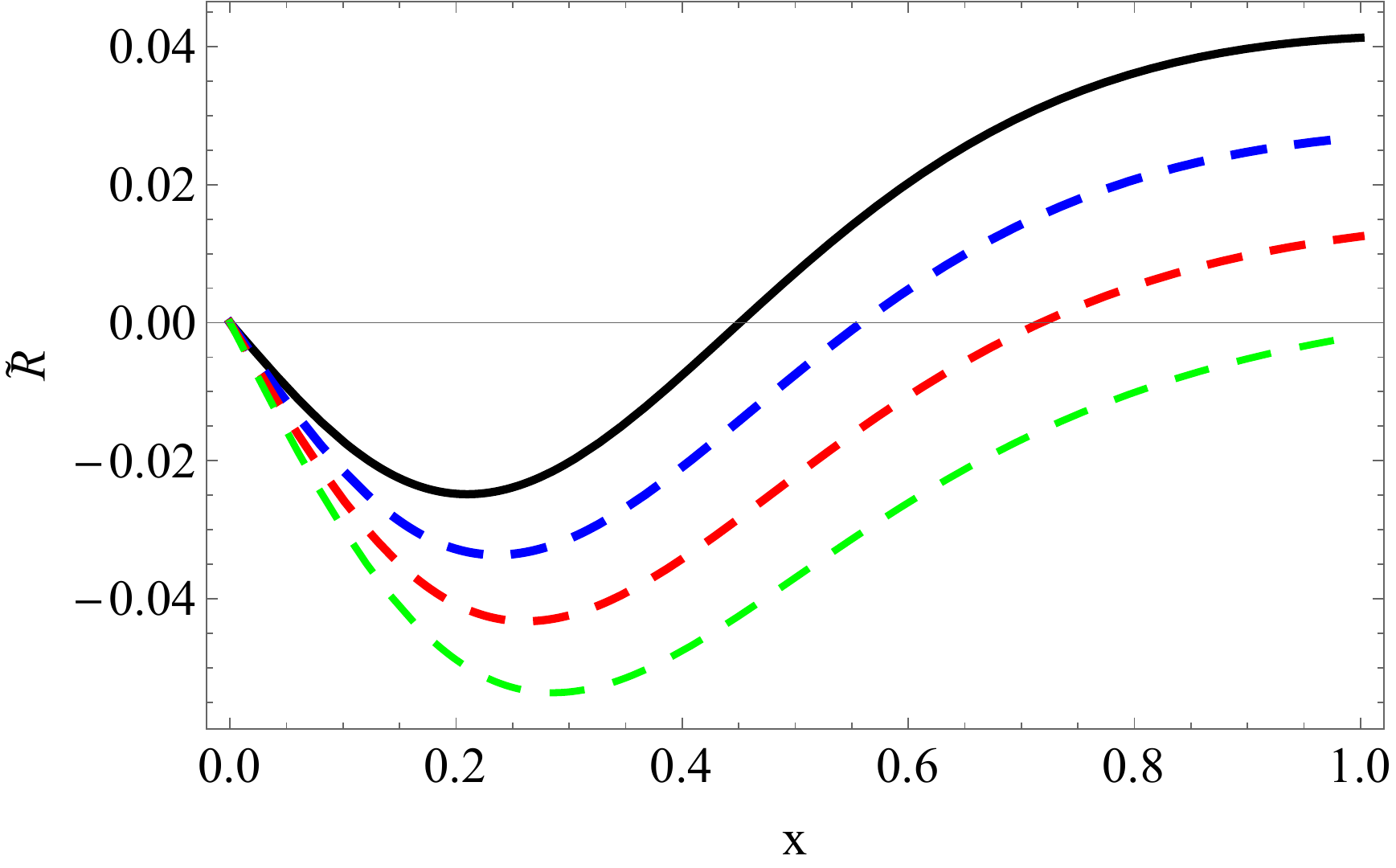}
\caption{\label{r5}
$\tilde{\mathcal{R}}$ as function of $x$, for 
$\beta=1$, $\Xi=0.3$ and $\Gamma=0.1$ (black line), $\Gamma=0.3$ (blue line) and $\Gamma=0.5$ (red line) and $\Gamma=0.7$ (green line)
}
\end{figure}

Following the same strategy, we now proceed to perturb the matter sector through variations of the parameters $\{\beta,\Xi\}$, so the system no longer satisfies the generalized hydrostatic equilibrium, TOV, equation (\ref{TOV}) and then the total force (\ref{fuerza-radial}) becomes different from zero,
\begin{eqnarray}
&&\tilde{\mathcal{R}}(\beta+\delta\beta,\Xi+\delta\Xi,x)=
\tilde{\mathcal{R}}(\beta,\Xi,x)+
\frac{\partial\tilde{\mathcal{R}}}{\partial\tilde{\beta}}\bigg|_{\beta,\Xi}\delta\beta\nonumber\\
&&\hspace{2cm}+\frac{\partial\tilde{\mathcal{R}}}{\partial\tilde{\Xi}}\bigg|_{\beta,\Xi}\delta\Xi+\mathcal{O}(\delta\Xi^{2},\delta\beta^{2}),
\end{eqnarray}
where we have defined $\Gamma$ analogously to (\ref{gamma1}) and (\ref{gamma2}), 
\begin{eqnarray}\label{gamma3}
\Gamma=\frac{\partial\tilde{\mathcal{R}}/\partial\tilde{\Xi}|_{\beta,\Xi}}{\partial\tilde{\mathcal{R}}/\partial\tilde{\beta}|_{\beta,\Xi}}.
\end{eqnarray}
Now, in the same way as before, we proceed to plot $\tilde{\mathcal{R}}$ as a function of $x$, for different values of $\beta$, $\Gamma$ and $\Xi$. In fig. \ref{r4} we show $\tilde{\mathcal{R}}$ as a function of $x$ for $\beta=1$, $\Gamma=0.1$ and various values for $\Xi$.
Note that as we approach to the Tolman IV solution ($\Xi=1$) the system becomes more stable. Also, the cracking point moves to outermost points of the stellar object. In fig. \ref{r5} we show $\tilde{\mathcal{R}}$ as a function of $x$ for $\beta=1$, $\Xi=0.3$ and various values for $\Gamma$. We observe that if the value of gamma increases the fracture moves towards more external regions of the compact object.

\begin{figure}[ht!]
\centering
\includegraphics[scale=0.5]{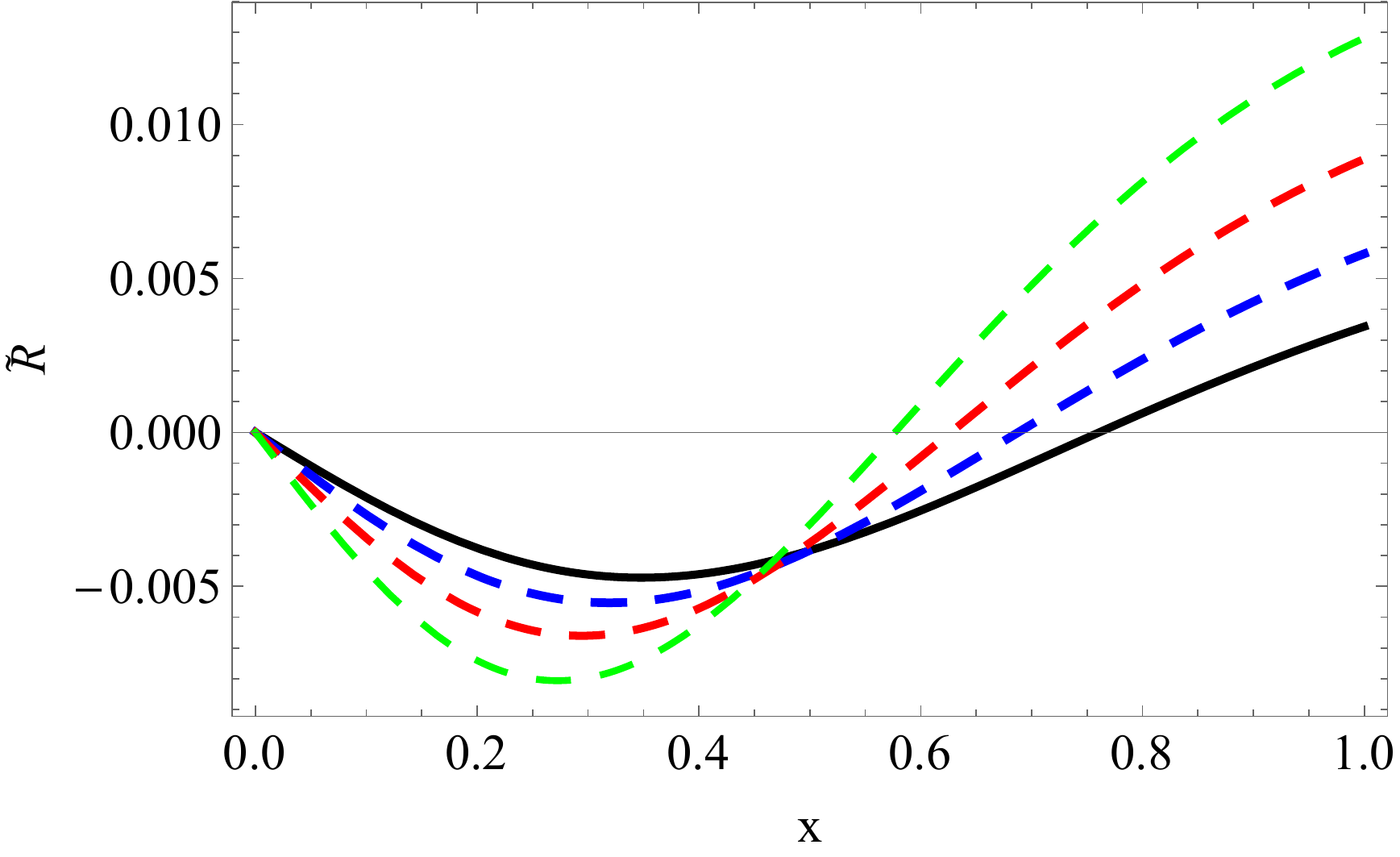}
\caption{\label{r6}
$\tilde{\mathcal{R}}$ as function of $x$, for 
$\Gamma=0.1$, $\Xi=0.3$ and $\beta=1.6,\ (u\approx 0.18-$black line), $\beta=1.5,\ (u\approx0.19-$blue line), $\beta=1.4, \ (u\approx0.2-$red line) and $\beta=1.3,\ (u\approx 0.21-$green line)
}
\end{figure}

In figure \ref{r6} we show 
$\tilde{\mathcal{R}}$ as a function of $x$ for $\Gamma=0.1$, $\Xi=0.3$ and various values of $\beta$, we observe that for more compact configurations the critical cracking point moves inward in the fluid distribution as expected.

\subsection{Anisotropic solutions with vanishing complexity}\label{model3}

\begin{figure}[ht!]
\centering
\includegraphics[scale=0.5]{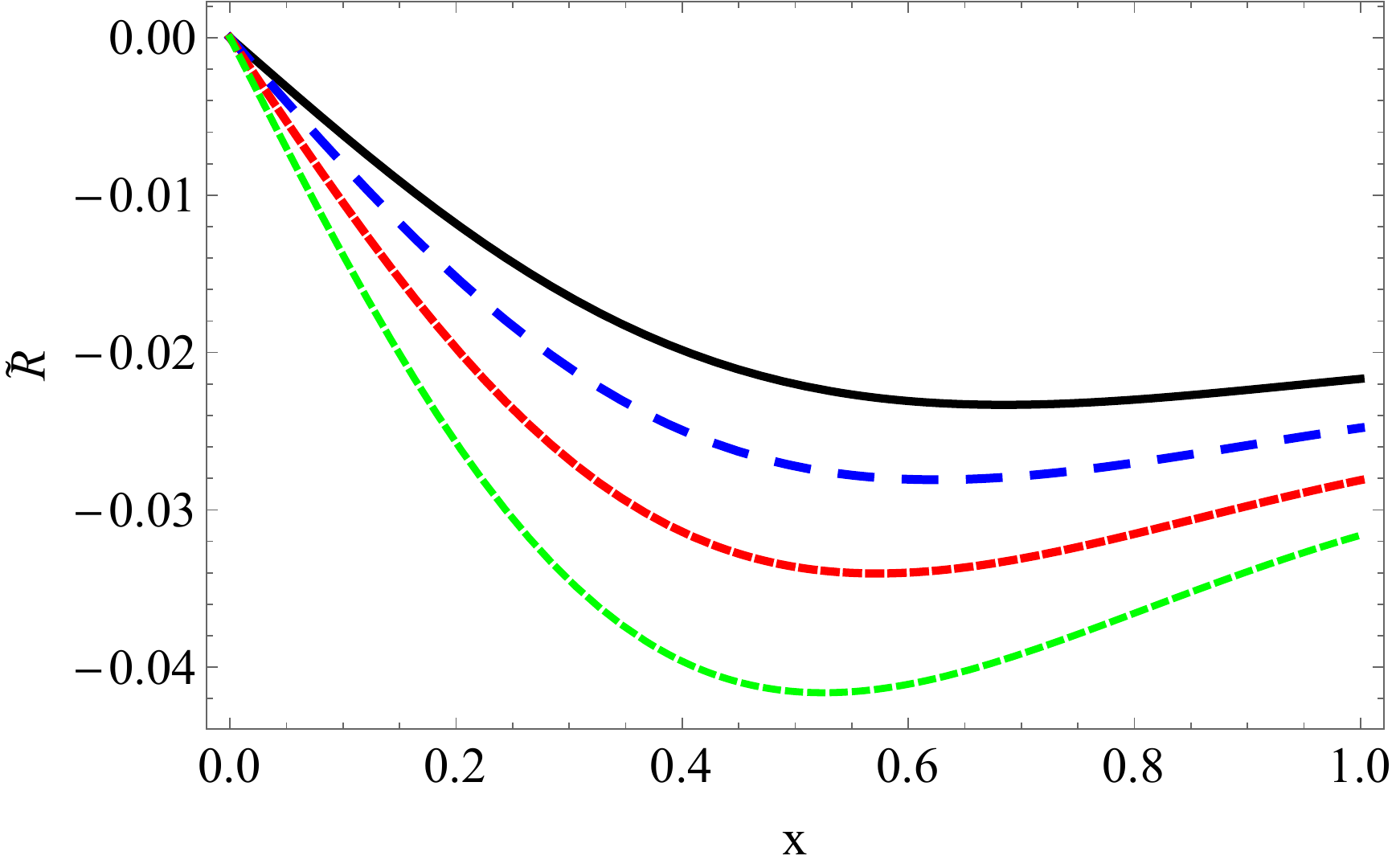}
\caption{\label{r7}
$\tilde{\mathcal{R}}$ as function of $x$, for 
$\beta=1.6,\ (u\approx 0.18-$black line), $\beta=1.5,\ (u\approx0.19-$blue line), $\beta=1.4, \ (u\approx0.2-$red line) and $\beta=1.3,\ (u\approx 0.21-$green line)
}
\end{figure}

In this section we study the cracking  for solutions with vanishing complexity. More precisely, the final solution is such that $Y_{TF}^{(m)}=-Y_{TF}^{\theta}$. Again, the seed solution is assumed as the Tolman IV model and the $\theta$--sector is obtained by Eq. (\ref{f2}). The result reads
\begin{eqnarray}
e^{\nu}&=&B^{2}
\left(
1+\frac{r^{2}}{A^{2}}
\right)\\
e^{-\lambda}&=&\frac{(A^{2}+r^{2})(2A^{2}-3r^{2}+6R^{2})}{(2A^{2}+3r^{2})(A^{2}+3R^{2})}\\
\rho&=&\frac{3(8A^{4}+2A^{2}(7r^{2}+3R^{2})+3r^{2}(3r^{2}+R^{2}))}{8\pi(2A^{2}+3r^{2})^{2}(A^{2}+3R^{2})}\nonumber\\
p_{r}&=&\frac{9(R^{2}-r^{2})}{8\pi(2A^{2}+3r^{2})(A^{2}+3R^{2})}\\
m&=&\frac{r^3 \left(4 A^2+3 \left(r^2+R^2\right)\right)}{2 \left(2 A^2+3 r^2\right) \left(A^2+3 R^2\right)}\\
\Pi&=&-\frac{3r^{2}(2A^{2}+3R^{2})}{8\pi(2A^{2}+3r^{2})^{2}(A^{2}+3R^{2})}.
\end{eqnarray}
Next, we can write
\begin{eqnarray}
\hat{p}_{r}&=&f(x)\\
\hat{\rho}&=&\frac{3 \left(8 \beta ^4+6 \beta ^2+\left(14 \beta ^2+3\right) x^2+9 x^4\right)}{8 \pi R^2  \left(\beta ^2+3\right)  \left(2 \beta ^2+3 x^2\right)^2}\\
\hat{m}&=&\frac{R x^3 \left(4 \beta ^2+3 x^2+3\right)}{2 \left(\beta ^2+3\right) \left(2 \beta ^2+3 x^2\right)}\\
\hat{\Pi}&=&-\frac{3  x^2 \left(2 \beta ^2 +3 \right)}{8 \pi R^{2} \left(\beta ^2 +3 \right) \left(2 \beta ^2 +3  x^2\right)^2},
\end{eqnarray}
with
\begin{eqnarray}
f(x)=-\frac{9 \left(x^2-1\right)}{8 \pi R^2 \left(\beta ^2+3\right)  \left(2 \beta ^2+3 x^2\right)}
\end{eqnarray}
Note that for this case, the only parameter we can perturb is $\beta$. In this sense, after perturbation, the total force reads
\begin{eqnarray}
\tilde{\mathcal{R}}(\beta+\delta\beta,x)=
\tilde{\mathcal{R}}(\beta,x)+
\frac{\partial\tilde{\mathcal{R}}}{\partial\tilde{\beta}}\bigg|_{\beta}\delta\beta+\mathcal{O}(\delta\beta^{2}).
\end{eqnarray}
In figure \ref{r7} we show the perturbed total force as a function of $x$ for various values of $\beta$.
It is noticeable that, in contrast with the other models considered here, there is no gravitational cracking. Indeed, the appearances of cracking is related to the roots of the polynomial
\begin{eqnarray}
&&32 \beta ^8+152 \beta ^6+288 \beta ^4+252 \beta ^2-27 x^8+135 x^6\nonumber\\
&&+\left(48 \beta ^4+144 \beta ^2-162\right) x^4\nonumber\\
&&+\left(48 \beta ^6+120 \beta ^4+54 \beta ^2+216\right) x^2,
\end{eqnarray}
which has no root for the allowed values of $\beta$ in accordance with the bound of the compactness parameter $u\le \frac{1}{3}$ given by Eq.(\ref{restbeta}).

\section{Conclusions}\label{concl}
Possible speculations of the occurrence of cracking in astrophysical settings, affecting the subsequent evolution of the system, have been invoked years ago \cite{VVCracking}. Situations like the collapse of a supermassive star where the cracking of a inner core would certainly change (probably enhance) the conditions for the ejection of the outer mantle in a supernova event (for both the prompt and the long thermo mechanism) \cite{SN1,SN2,SN3}. Although some works have addressed some connection showing that possible astrophysical phenomena, related to neutron stars, could be explained by the appearance of cracking inside these compact objects (see \cite{Ruderman1,Ruderman2,Ruderman3,Ruderman} for details), so far there is no concrete evidence and certainly there are clear differences between the models that we propose and those possible implications. We would like to emphasize that our aim here is not to model in detail any physical scenario but to call attention to the occurrence of cracking and its relationship with families of solutions with the same complexity parameter (also vanishing of the complexity parameter) analyzed using anisotropic solutions obtained in the framework of gravitational decoupling.

It is important to stress that the occurrence of cracking, has direct implications on the structure and evolution of the compact object, only at time scales that are smaller than, or at most, equal to, the hydrostatic time scale. What we do is to take a “snapshot” just after the system leaves the equilibrium. To find out whether or not the system will return to the state of equilibrium afterward would require an integration of the evolution equations.
However, it is clear that the occurrence of cracking would drastically affect the future structure and evolution of the compact object.\\

In this work, we obtained that both the anisotropy of the source and the complexity factor of the system play an important role in the appearance of cracking in all the models under consideration through the parameters involved: compactness of the source, decoupling parameter, perturbation ratio (that measure the difference between the variation of the parameters) and radius ratio (that measure the deviation between the radius of solutions with the same complexity). For example, in models 1 and 2 (section \ref{model1} and \ref{model2}, respectively) the cracking occurs for all the values of the perturbation ratio. However, as the value of this ratio increases, the cracking occurs in deeper regions of the fluid distribution for model 1 and in outer regions in model 2. Regarding to the compactness parameter, we showed that for the first model there is cracking for the more compact configurations while it is present for all the values in model 2. Furthermore, the size of the surface where cracking occurs increases as  the compactness takes greater values for model 1 and decreases for model 2. We also observed certain behaviours which are exclusive for the models separately. For example, in model 1 the decoupling parameter leads to a kind of screening effect on the anisotropy of the source 
which affect the occurrence of cracking. Indeed, the cracking is absent in the anisotropic solution and starts appearing as the screening effect becomes important
until the completely screened solution is reached (the isotropic model). The other example is exclusive to model 2 and corresponds to the effect of the radius ratio. This parameter measure how the total radius of any solution deviates from this of the
Tolman IV with the aim to maintain the same complexity. It was found that, as such a parameter decreases, the cracking, which is absent in the Tolman IV solution, occurs in deeper regions of the fluid. \\

As we have already mentioned, one of the purposes of this work was to relate cracking, not only with the anisotropy, which is already clear, but with a recent definition of the complexity factor given in \cite{complex1}. For a static fluid distribution, the simplest system is represented by a homogeneous energy density, and locally isotropic (in pressure) fluid. So it is assigned a zero value of the complexity factor for such a distribution. Also, it was shown that the “active” gravitational mass (Tolman mass) can be expressed through its value for the zero complexity case plus two terms depending on the energy density inhomogeneity and pressure anisotropy, respectively (for an arbitrary fluid distribution). These last terms may be expressed through the same scalar function ($Y_{TF}$) that defines the complexity factor. When the fluid is homogeneous in the energy density, and isotropic in pressure, this factor obviously vanishes, but also may vanish when the two terms containing density inhomogeneity and anisotropic pressure, cancel each other \cite{complex1}. This allows to have several different gravitational systems that can be characterized by the same (null) complexity: vanishing complexity may correspond to very different systems. 

Having highlighted this let's analyze our last model. The third model (section \ref{model3}) corresponds to a solution with vanishing complexity which only depends on the compactness of the source. In this case, cracking does not occur for any values of the compactness parameter. This represents a very interesting fact because it is known that in the cases where the initial (unperturbed) configuration consists of a locally isotropic fluid (perfect fluid) no cracking occurs and precisely these systems would be at the same level of complexity (the lowest) as the one discussed. Very different systems, representing compact objects, that share only the fact that their complexity factor vanishes, apparently, share also similar aspects in their (possible) future structure and evolution, at least as far as cracking is concerned. 

Finally, if the fluid configuration, used to model the stellar object, is homogeneous and isotropic (zero complexity factor) the appearance of this type of phenomena deserves particular attention. As shown in \cite{VVCracking} small deviations from local isotropy may lead to the occurrence of cracking. This implies that the subsequent evolution of the object, if such deviations are taken into account, may be very different from the situation where absolute local isotropy is assumed all along the evolution.

%%%%%%%%%%%%%%%%%%%%%%%%%%%
%\section*{Acknowledgements}
%%%%%%%%%%%%%%%%%%%%%%%%%%

%%%%%%%%%%%%%%%%%%%%%%%%%%%%%%%%%%%%%%%%%%%%%%%%%%%%%%%%%%%%%%%%%%%%%%%%%%%%%%%%%%%%%%%%%%%%%%%%

\end{document}